\documentclass[preprint]{ptephy_v1}

\preprintnumber{KYUSHU-HET-235} 

\usepackage{amsmath} 
\usepackage{amsthm} 
\usepackage{url} 

\usepackage{stmaryrd}

\DeclareMathOperator{\tr}{tr}

\newcommand{\Slash}[1]{{\ooalign{\hfil/\hfil\crcr$#1$}}}
\numberwithin{equation}{section}
\allowdisplaybreaks

\begin{document}

\title{Axial anomaly\\ in the gradient flow exact renormalization group}


\author{Yuki Miyakawa}
\affil{Department of Physics, Kyushu University, 744 Motooka, Nishi-ku,
Fukuoka 819-0395, Japan}





\begin{abstract}%
The gradient flow exact renormalization group (GFERG) is a formulation of
the exact renormalization group that keeps exact gauge invariance.
GFERG can keep also modified chiral invariance.
We will show that this formulation reproduces the correct axial anomaly in four dimensions.
\end{abstract}

\subjectindex{B05, B31, B32}

\maketitle

\section{Introduction}
\label{sec:1}
The exact renormalization group~(ERG)~\cite{Wilson:1973jj,Wegner:1972my,Wegner:1972ih}
(for reviews,~Refs.~\cite{Pawlowski:2005xe,Rosten:2010vm,Morris:1993qb,Dupuis:2020fhh}) gives the framework for non-perturbative quantum field theory.
The momentum cutoff used in the conventional derivation of ERG~\cite{Polchinski:1983gv}
breaks the gauge invariance,
which is a basic principle in particle physics.
To realize the gauge invariance,
a manifestly gauge-invariant ERG formulation has been
developed in~Refs.~\cite{Morris:1999px,Morris:2000fs,Arnone:2005fb,Morris:1998kz,Wetterich:2016ewc,Wetterich:2017aoy,Morris:2006in}.
Recently,
as one of ERG formulations
that can preserve the gauge invariance exactly,
the gradient flow exact renormalization group~(GFERG)
has been proposed in pure Yang-Mills theory~\cite{Sonoda:2020vut}.
GFERG for the gauge theory containing fermions, then, has been discussed in~Refs.~\cite{Miyakawa:2021hcx,Miyakawa:2021wus,Sonoda:2022fmk}.
The Wilson action that satisfies the ERG equation in scalar field theory
can be represented by the field obeying a diffusion equation.
GFERG is a new framework which uses a gauge covariant diffusion equation
(called the gradient flow equation) as the diffusion equation; hence it can keep
a manifest gauge invariance.
It is very important to confirm that GFERG is consistent with known phenomena in quantum field theory.
In this paper, as one of consistency checks, 
I will check whether GFERG in quantum electrodynamics~(QED) reproduces the U(1) axial anomaly or not.
In the context of the conventional ERG, the U(1) axial anomaly has been calculated in Refs.~\cite{Pernici:1997ie,Bonini:1997yv,Bonini:1994xj}

We work in the $D$-dimensional Euclidean space
although we set $D=4$ in the calculation of the axial anomaly. We use
the shorthand notation for the momentum integrals:
\begin{equation}
   \int_p\equiv\int\frac{d^Dp}{(2\pi)^D}.
\label{eq:(1.1)}
\end{equation}


\section{Perturbative solution of GFERG equation in QED}
\label{sec:2}
In order to calculate the axial anomaly in $D=4$ dimensions,
it is sufficient to know the Wilson action for the massless fermion up to the second-order in the gauge coupling~$e$.
On~$\mathcal{S}_\infty$~\cite{Wilson:1973jj}, which is the hyper-surface where all irrelevant parameters vanish, we expand the Wilson action in powers of~$e$,
\begin{equation}
   S=S^{(0)}+eS^{(1)}+e^2S^{(2)}+\dotsb,
 \label{eq:(2.1)}
\end{equation}
and we know the Wilson action which satisfies the GFERG equation in QED
up to $S^{(2)}$~\cite{Miyakawa:2021hcx,Miyakawa:2021wus}.
The zero-order action is given by
\begin{align}
   S^{(0)}
   &=-\frac{1}{2}\int_k\,
   A_\mu(k)A_\nu(-k)
   \left[
   \left(\delta_{\mu\nu}-\frac{k_\mu k_\nu}{k^2}\right)\frac{k^2}{e^{-2k^2}+k^2}
   +\frac{k_\mu k_\nu}{k^2}\frac{k^2}{\xi e^{-2k^2}+k^2}
   \right]
\notag\\
   &\qquad{}
   -\int_k\,\Bar{c}(-k)c(k)\frac{k^2}{e^{-2k^2}+k^2}
   -\int_p\,\Bar{\psi}(-p)
   \frac{\Slash{p}}{e^{-2p^2}+i\Slash{p}}\psi(p),
\label{eq:(2.2)}
\end{align}
where $\xi$ is the gauge fixing parameter. The first-order action is given by
\begin{equation}
  S^{(1)}=\int_{p,k}\,\Bar{\psi}(-p-k)e^{-(p+k)^2}\frac{1}{i}h_F(p+k)\widetilde{V}_\mu(p,k)
  e^{-p^2}\frac{1}{i}h_F(p)\psi(p)e^{-k^2}h_{\mu\rho}(k)A_\rho(k),
\label{eq:(2.3)}
\end{equation}
where
\begin{align}
  \widetilde{V}_\mu (p,k)
  =\gamma_\mu
  +2(\Slash{p}+\Slash{k})&p_\mu F((p+k)^2-p^2-k^2)
  +2\Slash{p}(p+k)_\mu F(p^2-(p+k)^2-k^2),
\label{eq:(2.4)}
\end{align}
and
\begin{align}
   \qquad\qquad{}h_{\mu\nu}(k)
   &\equiv
   \left(\delta_{\mu\nu}-\frac{k_\mu k_\nu}{k^2}\right)\frac{1}{e^{-2k^2}+k^2}
   +\frac{k_\mu k_\nu}{k^2}\frac{\xi}{\xi e^{-2k^2}+k^2},
   \\
   h_F(p)&\equiv
   \frac{i}{e^{-2p^2}+i\Slash{p}},\\
   F(x)&\equiv\frac{e^x-1}{x}.
\end{align}
The $\Bar{\psi}AA\psi$ term in the second-order action is given by
\begin{align}
   S^{(2)}|_{\Bar{\psi}AA\psi}
   &=\int_{p,k,l}\Bar{\psi}(-p-k-l)e^{-(p+k+l)^2}\frac{1}{i}h_F(p+k+l)\nonumber\\
   &\qquad \times \widetilde{V}_{\mu\nu}(p,k,l)
   e^{-p^2}\frac{1}{i}h_F(p)\psi(p)
   e^{-k^2}h_{\mu\rho}(k)A_\rho(k)e^{-l^2}h_{\nu\sigma}(l)A_\sigma(l),
 \label{eq:(2.8)}
\end{align}
where
\begin{align}
   \widetilde{V}_{\mu\nu}(p,k,l)
   &=\widetilde{V}_\mu(p+l,k)h_F(p+l)\widetilde{V}_\nu(p,l)
   +(\text{terms linear in $\gamma$ matrixes}).
\end{align}

We note that the function~$h_F(p)$ satisfies
\begin{align}
  h_F(p)+h_F(-p)=-2ih_F(p)h_F(-p)e^{-2p^2},
  \label{eq:(2.10)}
\end{align}
which is useful for the calculation in the next section.


\section{Calculation of the axial anomaly}
\label{sec:3}
GFERG can keep the invariance
under the modified chiral transformation~\cite{Miyakawa:2021hcx} and
if the Wilson action~$S$ is invariant under the modified chiral transformation, it
satisfies
\begin{align}
   &\int d^Dx\, \left\{ S \frac{\overleftarrow{\delta}}{\delta \psi(x)}\gamma_5 \psi(x)+
   \Bar{\psi}(x)\gamma_5 \frac{\delta}{\delta \Bar{\psi}(x)}S +2i S \frac{\overleftarrow{\delta}}{\delta \psi(x)}
   \gamma_5 \frac{\delta}{\delta \Bar{\psi}(x)}S
   -2i \tr \left[ \gamma_5 \frac{\delta}{\delta \Bar{\psi}(x)}S \frac{\overleftarrow{\delta}}{\delta \psi(x)} \right] \right\}\nonumber\\
   &=0
\end{align}
When the Wilson action is bi-linear in fermion fields,
\begin{align}
   S=i \int d^Dx\, d^Dy\, \Bar{\psi}(x)D(x,y)\psi(y)+\cdots,
    \label{eq:(3.1)}
\end{align}
then, $D(x,y)$ satisfies the Ginsparg--Wilson relation~\cite{Ginsparg:1981bj,Igarashi:1999rm,Igarashi:2002ba,Luscher:1998pqa,Neuberger:1997fp,Hasenfratz:1998jp,Neuberger:1998wv,Hasenfratz:1997ft,Hasenfratz:1998ri},
\begin{align}
   \gamma_5 D(x,y)+D(x,y)\gamma_5-2\int d^Dz\, D(x,z)\gamma_5D(z,y)=0,
    \label{eq:(3.2)}
\end{align}
and we can actually confirm that the terms in~Eq.~\eqref{eq:(2.1)} which contain the fermion fields
fulfill the relation~Eq.~\eqref{eq:(3.2)} in each order of~$e$ by using~Eq.~\eqref{eq:(2.10)}.

Under the infinitesimal chiral transformation with the local parameter~$\alpha(x)$,
\begin{align}
   \psi(x) \to \psi(x)+i\alpha(x)\gamma_5 \psi(x),\qquad{}\Bar{\psi}(x) \to \Bar{\psi}(x)+\Bar{\psi}(x) i\alpha(x)\gamma_5,
    \label{eq:(3.3)}
\end{align}
the Wilson action~\eqref{eq:(3.1)} changes as
\begin{align}
   S&\to S -\int d^Dx\, d^Dy\, [\alpha(y)-\alpha(x)]\Bar{\psi}(x)D(x,y)\gamma_5\psi(y)\nonumber\\
   &\qquad -2 \int d^Dx\, d^Dy\, d^Dz\, \alpha(x)\Bar{\psi}(x)D(x,z)\gamma_5 D(z,y)\psi(y),
    \label{eq:(3.4)}
\end{align}
where we have used~Eq.~\eqref{eq:(3.2)}.
If the integration measure $[d\Bar{\psi}d\psi]$ is invariant under the chiral transformation,
we obtain the following equation from~Eq.~\eqref{eq:(3.4)},
\begin{align}
   &-\int d^Dx\, d^Dy\, [\alpha(y)-\alpha(x)]\langle \Bar{\psi}(x)D(x,y)\gamma_5\psi(y)\rangle_{S}\nonumber\\
   &\qquad=2\int d^Dx\, d^Dy\, d^Dz\, \alpha(x) \langle \Bar{\psi}(x)D(x,z)\gamma_5 D(z,y)\psi(y)\rangle_{S}.
    \label{eq:(3.5)}
\end{align}
We regard $A_\mu$ as a background field and then the right-hand side of~Eq.~\eqref{eq:(3.5)} can be written as,
\begin{align}
   &2\int d^Dx\, d^Dy\, d^Dz\, \alpha(x) \langle \Bar{\psi}(x)D(x,z)\gamma_5 D(z,y)\psi(y)\rangle_{S}\nonumber\\
   &\qquad=-2\int d^Dx\, d^Dy\, d^Dz\, \alpha(x) \tr \left[D(x,z)\gamma_5 D(z,y)\langle\psi(y)\Bar{\psi}(x)\rangle_{S}\right]\nonumber\\
   &\qquad=-2\int d^Dx\, \alpha(x) \tr \left[\gamma_5 i D(x,x)\right],
    \label{eq:(3.6)}
\end{align}
where we have used
$\int d^Dy\, D(z,y) \langle\psi(y)\Bar{\psi}(x)\rangle_{S}=i \delta^{(D)}(z-x)$.

In what follows,
we determine the form of the $D=4$ axial anomaly when the gauge field carries a small momentum
compared to the cutoff.
This is the situation in which we have a universal form of the axial anomaly.
For the parameterization of the Wilson action~\eqref{eq:(2.1)}, we find
\begin{align}
   D(x,x)&=\int_p\,\frac{i\Slash{p}}{e^{-2p^2}+i\Slash{p}}\nonumber\\
   &\qquad{}-i\int_{p,k}\,e^{ikx}e^{-(p+k)^2}\frac{1}{i}h_F(p+k)\widetilde{V}_\mu(p,k)
   e^{-p^2}\frac{1}{i}h_F(p)e^{-k^2}h_{\mu\rho}(k)A_\rho(k)\nonumber\\
   &\qquad{}-i\int_{p,k,l}e^{ikx}e^{ilx}e^{-(p+k+l)^2}\frac{1}{i}h_F(p+k+l)
   \widetilde{V}_{\mu\nu}(p,k,l)\nonumber\\
   &\qquad\qquad\qquad{}\times e^{-p^2}\frac{1}{i}h_F(p)
   e^{-k^2}h_{\mu\rho}(k)A_\rho(k)e^{-l^2}h_{\nu\sigma}(l)A_\sigma(l).
   \label{eq:(3.a)}
\end{align}
In the small momentum limit, noting the following expansion formula,
\begin{align}
   h_{\mu\nu}(k)= \delta_{\mu\nu}+O(k^2)
    \label{eq:(3.7)},
\end{align}
we substitute~Eq.~\eqref{eq:(3.a)} into $-2\tr [\gamma_5 i D(x,x)]$ in the last line of~Eq.~\eqref{eq:(3.6)} and obtain
\begin{align}
   &-2\tr \left[\gamma_5 i D(x,x)\right]\nonumber\\
   &=-2 e^2 \tr \left[\gamma_5 \int_{k} \int_{l} e^{ikx} A_\mu(k) e^{ilx} A_\nu(l)
   \int_{p} e^{-(p+k+l)^2}\frac{1}{i}h_F(p+k+l)\widetilde{V}_{\mu\nu}(p,k,l) e^{-p^2}\frac{1}{i}h_F(p)\right]\nonumber\\
   &\qquad\qquad\qquad+O(k^2\text{ or }l^2).
\end{align}
For convenience, we change the integral variable as~$p\to p-l$. Noting the following expansion formula,
\begin{align}
   \widetilde{V}_\mu (p,k)
  = \gamma_\mu
  +2(\Slash{p}+\Slash{k})p_\mu
  +2\Slash{p}(p+k)_\mu+O(k^2),
\end{align}
we obtain
\begin{align}
   &-2\tr \left[\gamma_5 i D(x,x)\right]\nonumber\\
   &=-2 e^2 \tr \left[\gamma_5 \int_{k} \int_{l} e^{ikx} A_\mu(k) e^{ilx} A_\nu(l)
   \int_{p} e^{-(p+k)^2}\frac{1}{i}h_F(p+k)(\gamma_\mu+2(\Slash{p}+\Slash{k})p_\mu
   +2\Slash{p}(p+k)_\mu) \right.\nonumber\\
   &\qquad{}\left.\times h_F(p)(\gamma_\nu+2(\Slash{p}-\Slash{l})p_\nu
   +2\Slash{p}(p-l)_\nu)
   e^{-(p-l)^2}\frac{1}{i}h_F(p-l)\right]+O(k^2\text{ or }l^2),
   \label{eq:(3.b)}
\end{align}
where we have neglected the last terms of~Eq.~\eqref{eq:(2.4)}
because they and $h_F(p)$ are linear in the $\gamma$~matrix at most and thus they cannot
contribute to the left-hand side of~Eq.~\eqref{eq:(3.b)}.
Then using the expansion formula:
\begin{subequations}
\begin{align}
   &e^{-(p+k)^2}\frac{1}{i}h_F(p+k)\nonumber\\
   &=\left( \frac{e^{-3p^2}}{e^{-4p^2}+p^2}
   -\frac{2p\cdot k\, e^{p^2}(-1 +e^{4p^2} +3p^2e^{4p^2})}{(1+p^2e^{4p^2})^2}+O(k^2) \right)\nonumber\\
   &\qquad{}-\left( \frac{e^{-p^2}}{e^{-4p^2}+p^2}
   -\frac{2p\cdot k\, e^{3p^2}(-3 +e^{4p^2} +p^2e^{4p^2})}{(1+p^2e^{4p^2})^2}+O(k^2) \right)i(\Slash{p}+\Slash{k}),
\end{align}
\end{subequations}
and noting
$\tr (\gamma_5 \gamma_\mu\gamma_\nu\gamma_\rho\gamma_\sigma)=-4\epsilon_{\mu\nu\rho\sigma}$,
(i.e., we set $\gamma_5\equiv -\gamma_0 \gamma_1 \gamma_2 \gamma_3$ and $\epsilon_{0123}\equiv 1$),
we obtain
\begin{align}
   -2\tr \left[\gamma_5 i D(x,x)\right]
   &=-2i e^2 \int_{k} \int_{l} e^{ikx} A_\mu(k) e^{ilx} A_\nu(l)
   \int_{p} \frac{(4+16p^2)e^{-4p^2}}{(e^{-4p^2}+p^2)^3}\epsilon_{a\mu b\nu}k_a l_b + O(k^2\text{ or }l^2).
    \label{eq:(3.9)}
\end{align}
Since the integral in~Eq.~\eqref{eq:(3.9)} can be calculated as follows,
\begin{align}
  \int_{0}^\infty dx\, \frac{(4x+16x^2)e^{-4x}}{(e^{-4x}+x)^3}&=-2\int_{0}^\infty dx\, \frac{d}{dx}\frac{2e^{4x}x+1}{(e^{4x}x+1)^2}=2,
  \label{eq:(3.10)}
\end{align}
we obtain
\begin{align}
   -2\tr \left[\gamma_5 i D(x,x)\right]
   =\frac{ie^2}{16\pi^2}\epsilon_{a\mu b\nu}F_{a\mu}(x) F_{b\nu}(x)+ O(k^2\text{ or }l^2).
\end{align}

On the other hand, in the small momentum limit, the left-hand side of~Eq.~\eqref{eq:(3.6)} reduces to
\begin{align}
   \int d^Dx\, \alpha(x) \partial_\mu \langle \Bar{\psi}(x) \gamma_\mu \gamma_5 \psi(x) \rangle
   \equiv \int d^Dx\, \alpha(x) \partial_\mu \langle j_{5\mu}(x) \rangle.
    \label{eq:(3.11)}
\end{align}
In this way, we have the axial anomaly in $D=4$ as
\begin{align}
   \partial_\mu\langle j_{5\mu}(x)\rangle=\frac{ie^2}{16\pi^2}\epsilon_{\mu\nu\rho\sigma}F_{\mu\nu}(x) F_{\rho\sigma}(x),
    \label{eq:(3.12)}
\end{align}
which is the correct result.


\section{Conclusion}
\label{sec:4}
In this paper we have calculated the $D=4$ axial anomaly in QED by using
the Wilson action up to second-order in~$e$ that satisfies GFERG equation.
This anomaly caused by the fact
that the terms of the Wilson action bilinear in fermion fields satisfies the Ginsparg--Wilson relation.
The axial anomaly we have derived in~Eq.~\eqref{eq:(3.12)} is the desired result.

\section*{Acknowledgments}
I would like to thank Hidenori Sonoda and Hiroshi Suzuki for
helpful discussions.

\appendix



%



\let\doi\relax








\bibliographystyle{ptephy}
\bibliography{D=4axial-anomaly}

\end{document}